\documentclass[twocolumn,nofootinbib]{revtex4-1}
\usepackage{latexsym,epsfig,amssymb, amsmath,nicefrac}
\usepackage{epsfig}

              \newcommand{\rf}[1]{(\ref{#1})}

\def\bfone{\relax{\rm 1\kern-.35em 1}}

\newcommand{\be}{\begin{equation}}
\newcommand{\ee}{\end{equation}}
\newcommand{\ben}{\begin{displaymath}}
\newcommand{\een}{\end{displaymath}}
\newcommand{\bea}{\begin{eqnarray}}
\newcommand{\eea}{\end{eqnarray}}

\newcommand{\bean}{\begin{eqnarray*}}
\newcommand{\eean}{\end{eqnarray*}}

\newcommand{\vp}{\varphi}

\def\K{K{\"a}hler}

\begin{document}

\title{\Large{Escher in the Sky}}

\author{Renata Kallosh and Andrei Linde}

\affiliation{Department of Physics and SITP, Stanford University, \\ 
Stanford, California 94305 USA, kallosh@stanford.edu, alinde@stanford.edu}

\begin{abstract}
We give a brief review of the history of inflationary theory and then concentrate on the recently discovered set of inflationary models called cosmological $\alpha$-attractors. These models provide an excellent fit to the latest observational data. 
 Their predictions $n_{s} \approx 1-2/N$ and $r \approx 12\alpha/N^{2}$ 
  are very robust with respect to the modifications of the inflaton potential. An intriguing interpretation of $\alpha$-attractors is based on a geometric moduli space with a boundary: a Poincar\'e disk model of a hyperbolic geometry with the radius $\sqrt{3\alpha}$,   
   beautifully represented by the Escher's picture Circle Limit IV.  In such models, the amplitude of the gravitational waves is proportional to the square of the radius of the Poincar\'e disk. 

\end{abstract}

\maketitle

\smallskip


\section{Introduction}\label{intro}

During  the last 35 years inflationary theory evolved from something that could look like a beautiful science fiction story to the well established scientific paradigm describing the origin of the universe and its large scale structure. Many of its predictions have been already confirmed by observational data, see e.g.  \cite{Ade:2015tva,Planck:2015xua}. And yet the development of this branch of science is not over. In this paper we will briefly remember the first steps of its development, and then relate them to a broad set of inflationary models which seem to fit observational data particularly well, and which make predictions nearly independent on the shape of their inflationary potentials. We called these theories ``cosmological attractors.'' As we will show, this class of models is closely related to some of the pioneering inflationary models such as the simplest versions of the chaotic inflation scenario \cite{Linde:1983gd,Linde:2005ht} and  the Starobinsky model \cite{Starobinsky:1980te}. But what makes these theories especially interesting is their geometric nature and supergravity realization, bringing us back to the discussion of the Poincar\'e disk and Escher's paintings. To put these theories into proper context, we will remind here some basic facts from the history of development of inflationary models.

\section{A brief history of inflationary ideas}

The development of inflationary cosmology had more than a fair share of twists and turns, and it is very different from its first implementations. It took several years until the contours of this theory became sufficiently well established.

The first model of inflationary type was proposed by Starobinsky \cite{Starobinsky:1980te}. In its original form, it was based on adding the contribution of conformal anomaly to the Einstein theory, which required existence of an enormous variety of different types of elementary particles contributing to the anomaly.  Instead of attempting to solve the homogeneity and isotropy problems, which is the defining feature of all inflationary models, Starobinsky assumed that  the universe was homogeneous and isotropic from the very beginning, and emphasized that his scenario was ``the extreme opposite of Misner's initial chaos''   \cite{Starobinsky:1980te}. The goal of the model was to solve the singularity problem by starting the evolution in a non-singular de Sitter state. However, dS state in his scenario was unstable, with a finite decay time \cite{Mukhanov:1981xt}, and therefore it could not exist at $t \to -\infty$. 

The main goals of inflationary theory were formulated for the first time in the context of old and new inflation  \cite{Guth:1980zm,Linde:1981mu,Albrecht:1982wi}. These models were based on an assumption that the universe initially was in a state of thermal equilibrium at an extremely high temperature, 
and then it supercooled and inflated in a state close to the top of the potential $V(\phi)$ At that time, this assumption seemed established beyond any reasonable doubt. However, old inflation did not quite work, as pointed out by its author \cite{Guth:1980zm}, and it did not lead to perturbations of the cosmic microwave background radiation, which were predicted in \cite{Mukhanov:1981xt,Starobinsky:1982ee} and discovered by COBE, WMAP and Planck. New inflation resolved most of the problems of old inflation, but it was also ruled out a year later, for many reasons discussed in \cite{Linde:2005ht}. After the first successes of inflationary theory, its future could appear quite bleak.  As Hawking said in his book back in 1988, ``the new inflationary model is now dead as a scientific theory, although a lot of people do not seem to have heard about its demise and write papers as if it were viable''  \cite{Hawking1988}.

The situation changed with the invention of the chaotic inflation scenario  \cite{Linde:1983gd}. It was proposed as an alternative to new inflation, after it was realized that the assumption of the hot Big Bang, high temperature phase transitions and supercooling did not help to formulate a successful inflationary theory. In fact, these basic assumptions,  the standard trademarks of old and new inflation, made inflation much more difficult to implement. If, instead, one simply considers the universe with different initial conditions in its different parts (or different universes with different values of fields in each of them), one finds that in many of them inflation may occur. It makes these parts exponentially large, thus producing exponentially large islands of order from the primordial chaos. Hence the name: chaotic inflation.

An important feature of this scenario is its versatility and the broad variety of models where it can be implemented. Examples of chaotic inflation models proposed in the 80's included models with monomial and polynomial potentials, and any other models where the slow roll regime was possible. This regime is possible in small field models, with the potentials of the new inflation type, or with models with the Higgs-like potential $\sim \lambda(\phi^{2}-v^{2})^{2}$ with $v \gg 1$  \cite{Linde:1984cd}. Models of that type later have been  called ``hilltop inflation'' \cite{Boubekeur:2005zm}. Another example was the first version of chaotic inflation with a plateau potential of the type $V \sim a(1-e^{-b|\phi|})$ \cite{Goncharov:1983mw}. It was, simultaneously, the first realization of chaotic inflation in supergravity, providing an excellent fit to the latest observational data  \cite{Ade:2015tva,Planck:2015xua}. 
In what follows we will call it the GL model. 

In 1983-1988, the Starobinsky model  \cite{Starobinsky:1980te} experienced significant modifications. First of all, it was reformulated as a theory $R + aR^{2}$, and initial conditions for inflation in this theory were formulated along the lines of the chaotic inflation scenario \cite{Barrow:1983rx,Starobinsky:1983zz,Kofman:1985aw}. This resolved the problem with initial conditions of the original version of this model  \cite{Starobinsky:1980te}. Then in 1988, using the results by Whitt \cite{whitt}, this model was reformulated in its present form, in terms of the scalar field theory with the plateau potential $V \sim a(1-e^{-b\phi})^{2}$  \cite{Barrow:1988xi}, very similar to potential of the GL model  \cite{Goncharov:1983mw}. 

Other popular versions of chaotic inflation where developed in the 90's. The authors of the ``natural inflation'' scenario said that ``our model is closest in spirit to chaotic inflation''  \cite{Freese:1990rb}. The hybrid inflation scenario  \cite{Linde:1991km}, was also introduced as a specific version of the chaotic inflation scenario.  Step by step, chaotic inflation replaced new inflation in its role of the main inflationary paradigm. Rather than describing some particular subset of inflationary models, it describes the most general approach to inflationary cosmology, which can easily incorporate ideas of quantum cosmology, eternal inflation, inflationary multiverse, and string theory landscape \cite{DeWitt:1967yk,Vilenkin:1982de,Linde:1984ir,Zeldovich:1984vk,Vilenkin:1984wp,Linde:2004nz,Vilenkin:1983xq,Linde:1986fd,Linde:1986fc,Linde:1993xx,Bousso:2000xa,Kachru:2003aw,Douglas:2003um,Susskind:2003kw}.

But this did not happen overnight. Chaotic inflation was so much different from old and new inflation that for a while it was psychologically difficult to accept. 
Even now, 30 years since the demise of old and new inflation, most of the college books on physics and astrophysics still describe inflation as exponential expansion in the false vacuum state during cosmological phase transitions with supercooling in Grand Unified Theories. That is why a significant part of the first book on inflation \cite{Linde:2005ht} was devoted to the discussion of new inflation versus chaotic inflation. 
 
By now, this discussion is over, most of the existing models of inflation are based on the main principles of chaotic inflation. However this introduced a purely terminological issue: every new inflationary model belonging to the general class of chaotic inflation is introduced with its own name. That is why some authors invented a different classification of models and  say, incorrectly, that chaotic inflation describes only models with monomial potentials, or only large field models, as opposite, e.g., to the hilltop inflation, natural inflation and hybrid inflation. In this paper we use the original  definition of chaotic inflation following \cite{Linde:1983gd,Linde:2005ht}.

\section{$\alpha$ attractors: T-models and E-models}\label{single}

Despite the generality of the chaotic inflation scenario described in the previous section, there is a good reason why in minds of many cosmologists chaotic inflation is often associated with the simplest model with a quadratic potential, with the Lagrangian \cite{Linde:1983gd,Linde:2005ht},
 \be
 {1\over \sqrt{-g}} \mathcal{L} = {1\over 2}   R - {1\over 2} {\partial \phi^2}  - {1\over 2}{m^2} \phi^2   \,  .
\label{quadratic}\ee
Nothing can be simpler than that, and yet it leads to inflation. This simplicity served as one of the main arguments in favor of naturalness of inflationary theory: No need for false vacuum states, complicated potentials and cosmological phase transitions with supercooling, just take a theory with a simple harmonic oscillator potential, put it into a cosmological background, and we are done.

However, the new observational data strongly suggest that this model predicts too large amplitude of gravitational waves, and therefore it requires  modifications   \cite{Ade:2015tva,Planck:2015xua}. The simplest modification, which we are going to discuss in this paper,  is provided by the models of cosmological $\alpha$ attractors  \cite{Kallosh:2013hoa,Ferrara:2013rsa,Kallosh:2013yoa,Cecotti:2014ipa,Galante:2014ifa,Linde:2014hfa,Kallosh:2015lwa}. For example, one may consider a theory with the Lagrangian 
 \be
 {1\over \sqrt{-g}} \mathcal{L}_{\rm T} = {1\over 2}   R - {1\over 2} {\partial \phi^2\over (1-{\phi^{2}\over 6\alpha})^{2}}  - {1\over 2}{m^2} \phi^2   \,  .
\label{cosmo}\ee
Here $\phi(x)$ is the scalar field, the inflaton. The parameter  $\alpha$  can take any positive value. The kinetic term of the inflaton is not canonical, which will play a very important role in what follows.
At $\alpha \rightarrow \infty$ this model coincides with the  simple chaotic inflation model \rf{quadratic}. 

The field $\phi$ is not canonically normalized. It must satisfy the condition $\phi^2<6\alpha$, so that the sign of the inflaton kinetic term is non-singular.  But one can easily go to canonically normalized variables $\vp$. For any finite $\alpha$ one can solve equation ${\partial \phi\over 1-{\phi^{2}\over 6\alpha}} = \partial\vp$, which yields
\be 
\phi = \sqrt {6 \alpha}\, \tanh{\varphi\over\sqrt {6 \alpha}} \ .
\ee
The boundary of the moduli space $\phi = \pm \sqrt {6 \alpha}$ becomes $\pm \infty$ in terms of the canonically normalized inflaton field $\vp$, and the quadratic potential becomes $V= 3\alpha m^{2} \tanh^2{\varphi\over\sqrt {6 \alpha}}$. 
We called such $\alpha$-attractors  `T-models': their potentials depend on $\tanh^2{\varphi\over\sqrt {6 \alpha}}$, they are symmetric with respect to the change $\vp\to -\vp$ and look like letter T  \cite{Kallosh:2013hoa}. All potentials $V(\phi^{2})$ belong to the general class of T-models, which includes the GL model  \cite{Goncharov:1983mw}, which was the first successful implementation of chaotic inflation in supergravity. In modern language, GL model described $\alpha$ attractor  with $\alpha= 1/9$ and the potential $V(\phi) \sim \phi^{2}(1-{3\over 8}\phi^{2})$ \cite{Linde:2014hfa,Kallosh:2015lwa}.
\begin{figure}[ht!]
\centering
{\hspace{5mm}
\includegraphics[scale=.43]{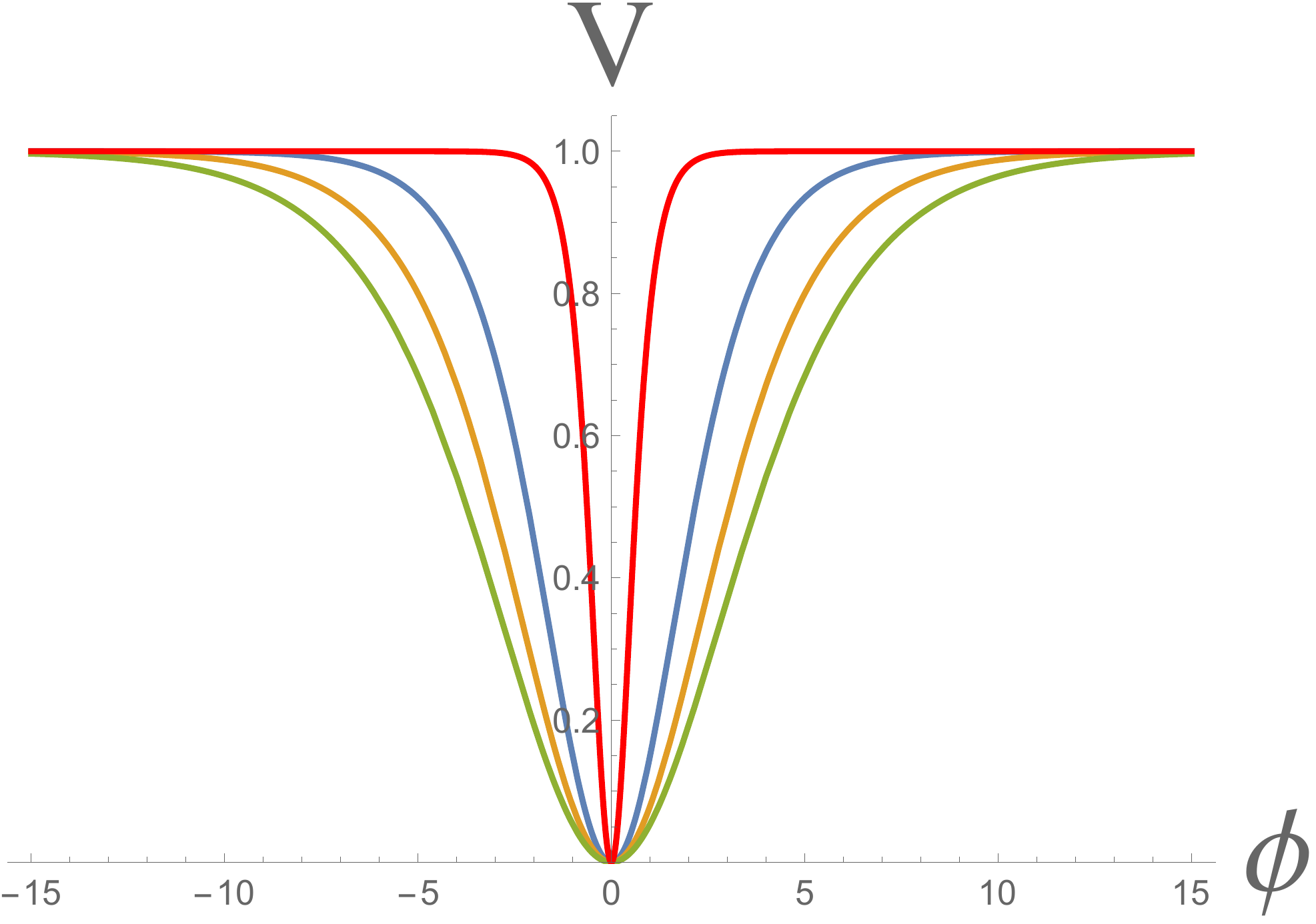}
\label{figT}}~~~~~
\caption{\footnotesize Blue, brown and green lines show the potentials of the T-models with $ V \sim \tanh^2{\varphi\over\sqrt {6 \alpha}}$ for $\alpha = 1, 2, 3$ correspondingly.  The red line in the center shows the potential of the GL model  \cite{Goncharov:1983mw}.  }
\end{figure}

\begin{figure}[ht!]
\begin{center}
\includegraphics[width=8.5cm]{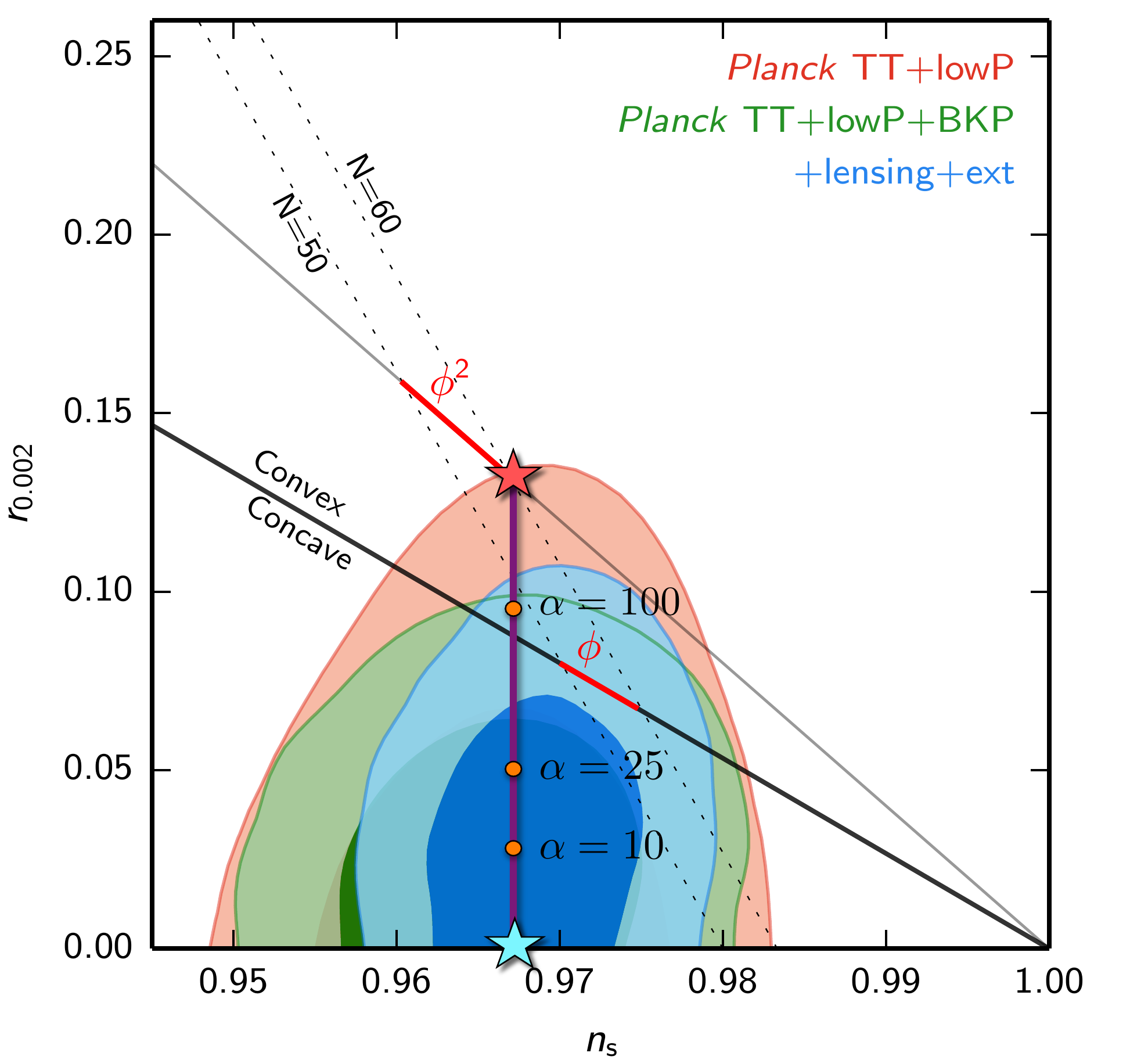}
\vspace*{-0.4cm}
\caption{\footnotesize Predictions of the simplest $\alpha$-attractor  T-model with the potential $V\sim  \tanh^2{\varphi\over\sqrt {6 \alpha}}$  for N = 60 cut through the most interesting part of the Planck 2015 plot for $n_{s}$ and $r$ \cite{Planck:2015xua}. }
\label{f1}
\end{center}
\vspace{-0.3cm}
\end{figure}

In the leading order in the inverse number of e-foldings $N$, for $\alpha \ll N$, the slow roll parameters $n_{s}$ and $r$ for T-models are
\be\label{nsr}
 1 -n_{s} = {2\over N}\, , \qquad r =  {12\alpha \over N^{2} } \ .
 \ee
For large $\alpha$, the prediction for $n_{s}$ practically does not change, but the growth of $r$ slows down: $r \approx  {12\alpha \over N(N+3\alpha/2) }$ \cite{Kallosh:2013yoa}.  The exact interpolating values of $n_{s}$ and $r$  for the theory $V = \tanh^2{\varphi\over\sqrt {6 \alpha}}$ are plotted in  Fig. \ref{f1} by a thick purple vertical line superimposed with the results for $n_s$ and $r$ from the Planck 2015 data release \cite{Planck:2015xua}. This line begins at the point corresponding to the predictions of the simplest quadratic model ${m^{2}\over 2}\phi^{2}$ for $\alpha > 10^{3}$ (red star), and then, for $\alpha\lesssim 40$, it enters the region most favored by the Planck data. For $\alpha = 1$, these models give the same prediction $r \sim 12/N^{2}$ as the Starobinsky model,  the Higgs inflation model \cite{Salopek:1988qh},  and the broad class of superconformal attractors \cite{Kallosh:2013hoa}. Then the same vertical line continues further down towards the prediction $r \sim 4/3N^{2}$ of the GL model \cite{Goncharov:1983mw,Linde:2014hfa} corresponding to $\alpha = 1/9$. Then it goes even further, all the way down to $r \to 0$ in the limit $\alpha \to 0$. Predictions of all models with $\alpha \lesssim O(1)$ are so close to each other, that they are covered by the same blue star in Fig. \ref{f1}.

One can show that in the large $N$ limit   {\it not only $n_{s}$, but also the amplitude of scalar perturbations in this class of models does not depend on $\alpha$}; it depends only on $N$ and $\mu$. For $N = 60$, this amplitude matches the Planck 2015 normalization if $\mu \approx 10^{{-5}}$. 

Moreover, for sufficiently small $\alpha \lesssim O(1)$, the predictions of $\alpha$-attractors in the large $N$ limit almost do not depend on whether we take the potential   $\tanh^2{\varphi\over\sqrt {6 \alpha}}$, or use a general class of potentials $V(\phi) =f^{2}(\phi) = f^{2}(\tanh{\varphi\over\sqrt {6 \alpha}})$ for a rather broad set of choices of the functions $f(\phi)$.  This stability of predictions, as well as their convergence  to one of the two attractor points shown in Fig. \ref{f1} by the red and blue stars, is the reason why we called these theories the cosmological attractors. The latest Planck 2015 result $n_{s} = 0.968\pm 0.006$ \cite{Planck:2015xua} almost exactly coincides with the prediction of the simplest T-models for $N = 60$. These properties of T-models are quite striking. Since their predictions can match any value of $r$  from $ 0.14$ to $0$, see Fig. \ref{f1}, these models may have lots of staying power.

\begin{figure}[ht!]
\centering
{
\includegraphics[scale=.42]{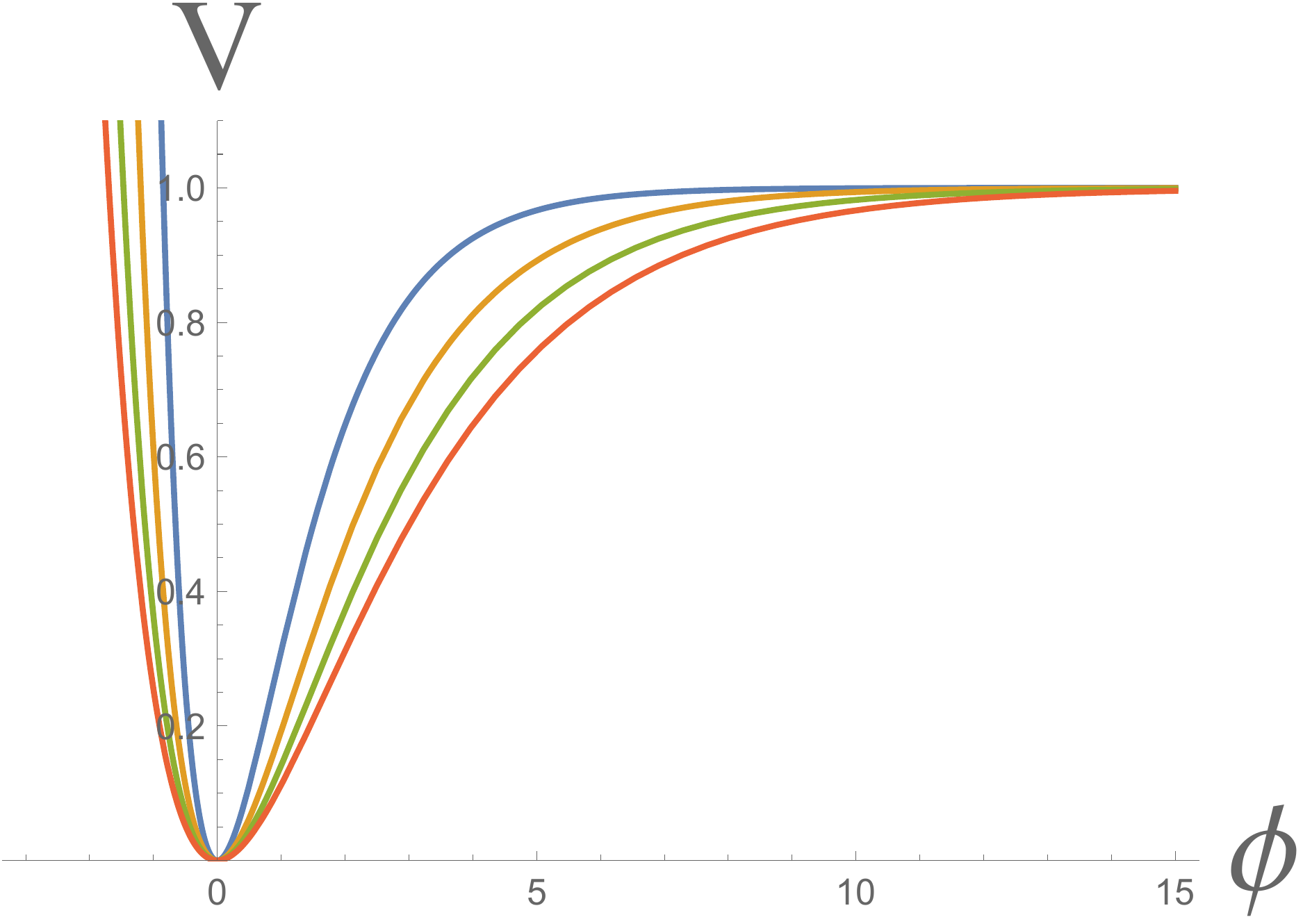}
\label{fig:newlongb}}
\caption{\footnotesize E-model potential  $\alpha\mu^{2}(1-e^{-{\sqrt {2\over 3 \alpha}}\varphi})^{2}$ in units of $\alpha\mu^{2} = 1$ for $\alpha = 1, 2, 3, 4$. Smaller $\alpha$ correspond to more narrow minima of the potentials. The blue line shows the potential of the Starobinsky model, which belongs to the class of E-models with $\alpha = 1$.}
\end{figure}

\begin{figure}[htb]
\hspace{1mm}\vspace{-5mm}
\begin{center}
\includegraphics[width=8.5cm]{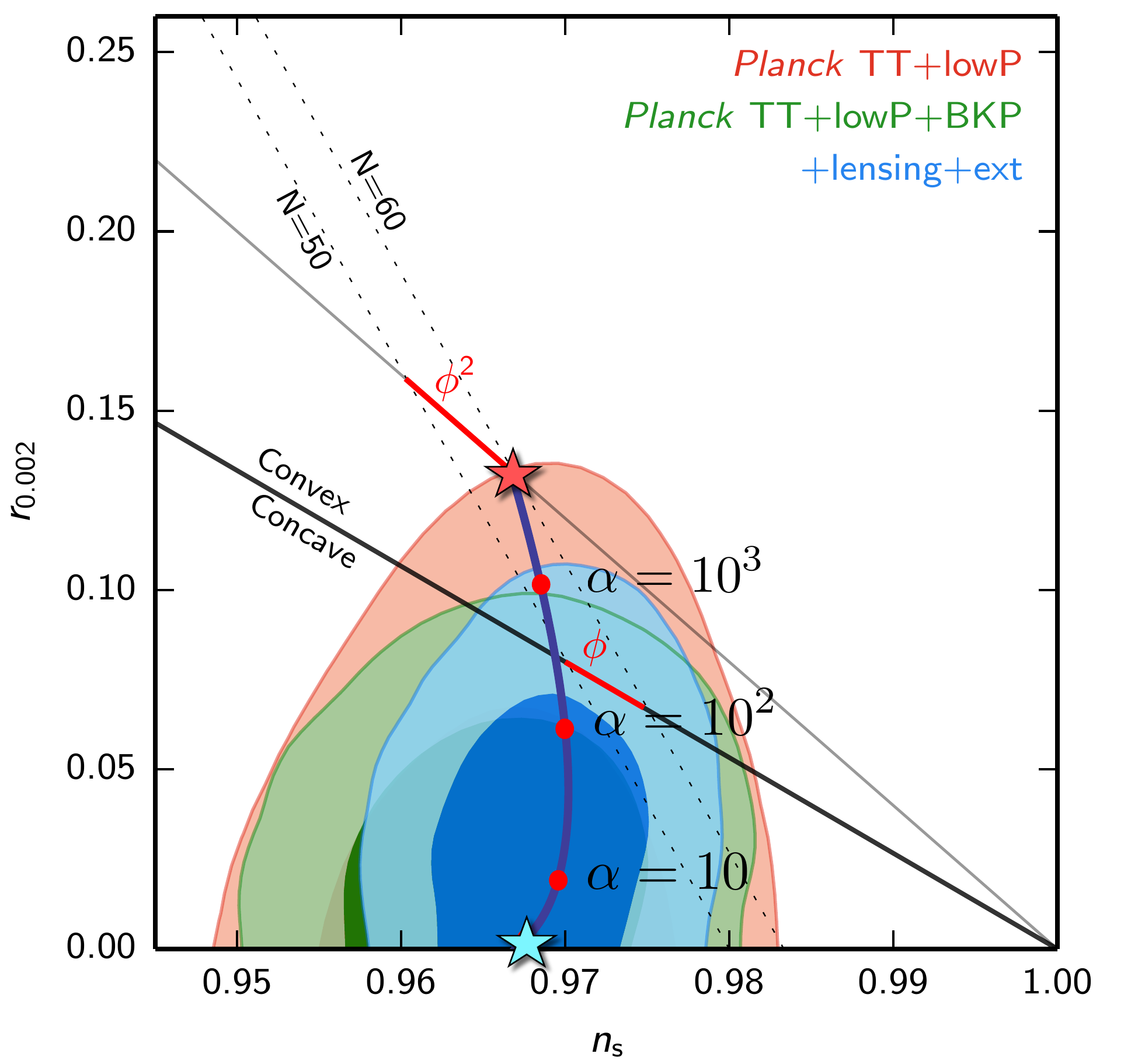}
\vspace*{-0.4cm}
\caption{\footnotesize Predictions of  E-models with $V\sim  (1-e^{-{\sqrt {2\over 3 \alpha}}\varphi})^{2}$. }
\label{f2}
\end{center}
\vspace{-0.4cm}
\end{figure}

As an example of a set of $\alpha$ attractors corresponding to a slightly more complicated choice of the function $f(\phi)$, we will describe now a set of models with  $V(\phi) =f^{2}(\phi) =\bigl({\phi\over 1+\phi/ \sqrt{6\alpha}}\bigr)^{2}$: 
 \be
 {1\over \sqrt{-g}} \mathcal{L}_{\rm E} = {1\over 2}   R - {1\over 2} {\partial \phi^2\over (1-{\phi^{2}\over 6\alpha})^{2}}  - {1\over 2}{m^2} {\phi^2\over (1+   {\phi\over \sqrt{6\alpha}} )^{2}}\,  .
\label{cosmoE}\ee
We called this set of $\alpha$ attractors `E-models' because the potential  of these models has an explicit exponential dependence on the canonically normalized field $\vp$, asymmetric with respect to the change $\vp \to
 -\vp$: \ $V~\sim~(1-e^{-{\sqrt {2\over 3 \alpha}}\varphi})^{2}$. In the special case $\alpha = 1$ this potential coincides with the potential of the Starobinsky-Whitt model \cite{Starobinsky:1983zz,Kofman:1985aw,whitt}, which represents this model as a member of the general class of $\alpha$-attractors, see Fig. \ref{fig:newlongb}. Predictions of these models are shown in Fig. \ref{f2}.
 
All of these models have the same kinetic term but different potentials. They have two common features. Generically,  they have two attractor points, shown by the red and blue stars in Figs. \ref{f1} and \ref{f2}, describing the limiting behavior for $\alpha\to \infty$ and $\alpha\to 0$. More importantly, for sufficiently small $\alpha$ (i.e. in the limit when the size of the moduli space becomes small) their cosmological predictions are very stable with respect to even very significant modifications  of the potentials.

This property was explained in  \cite{Kallosh:2013hoa,Ferrara:2013rsa,Kallosh:2013yoa}, and it was formulated in a particularly general way in \cite{Galante:2014ifa}: The kinetic term in this class of models, as well as in many other models of cosmological attractors, has a pole near the boundary of the moduli space. If inflation occurs in a vicinity of such a pole (which happens for sufficiently small $\alpha$), and the potential near the pole can be well represented by its value and its first derivative near the pole, all other details of the potential far away from the pole (from the boundary of the moduli space) become unimportant for making cosmological predictions. In particular, the  spectral index depends solely on the order of the pole, while the tensor-to-scalar ratio also involves the residue  \cite{Galante:2014ifa}. All the rest is practically irrelevant, as long as the field after inflation falls into a stable minimum of the potential with a tiny value of the vacuum energy and stays there.

From the point of view of constructing single field inflationary models, everything becomes nearly trivial: Take any model with a pole in the kinetic term and a potential which has a  minimum, and we are done, independently of many other details of the theory, in perfect agreement with observations. In this sense, everything becomes as transparent as in the simplest chaotic inflation model \rf{quadratic}, but more general and stable with respect  to the choice of the inflationary potential. One may argue that what this new class of models does for inflation is somewhat similar to what inflation did for cosmology. Inflation makes the structure of the observable part of the universe very stable with respect to the choice of initial conditions in the early universe. Meanwhile the cosmological attractors make inflationary predictions which are very stable with respect to the choice of the inflaton potential.

But can we implement this scenario in models related to advanced  theories of fundamental interactions? And if the properties of the kinetic term are so important, is it possible that this class of models may have some interesting interpretation in terms of geometry of the moduli space? The rest of the paper will be dedicated to the discussion of these issues, under the guidance of Poincar\'e and Escher, as well as of many of our friends in the supergravity/string theory community.

\section{The hyperbolic plane  ${ \mathbb{H}^2}$}  

In the previous sections we briefly described an interpretation of $\alpha$-attractors in simple phenomenological models of a single scalar field. However, the main goal of this paper is to show that the parameter $\alpha$ in advanced cosmological attractor models based on supergravity is best described by a size of the Poincar\'e disk famously represented by the Escher's  Circle Limit IV, see Fig. \ref{f2}. Namely, the radius square of the boundary of this circle $R^2$, in the context of our cosmological models,  is given by $3\alpha$. The smaller the level of primordial gravity waves, the smaller the circle! Current data implies that $R^2\lesssim 75$  for the simplest T-model and  $R^2\lesssim 300$ for the simplest E-model.

\begin{figure}[ht!]
\begin{center}
\includegraphics[width=7.2cm]{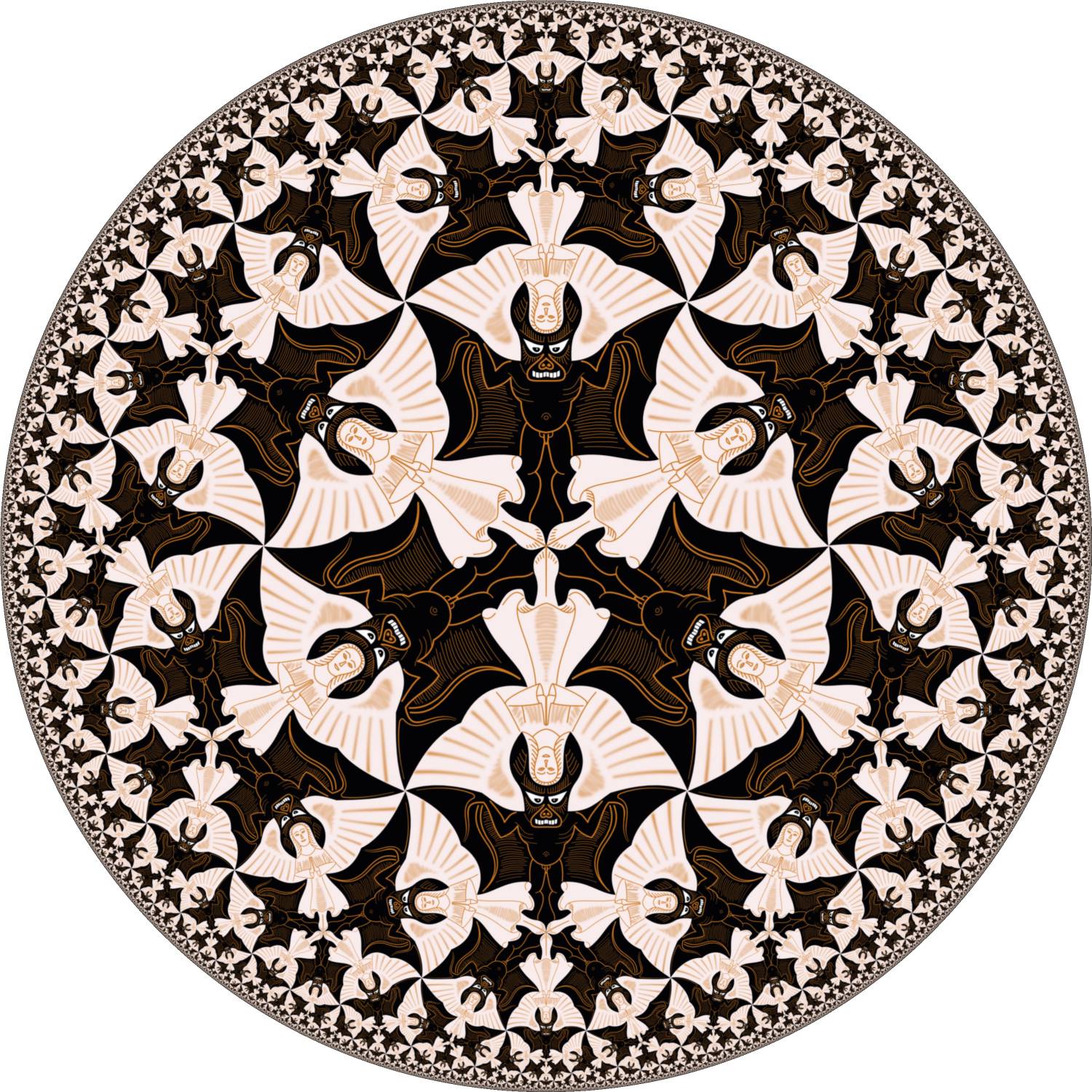}
\caption{\footnotesize A computer generated version of Escher's picture Circle Limit IV (Heaven and Hell) by V. Bulatov, http://bulatov.org/math/1201/.  It presents a Poincar\'e disk model of a hyperbolic geometry.  
The radius square of the disk in the context of our cosmological models  is $R^2=3\alpha$.
The curvature of this manifold ${\cal R}_{ \mathbb{H}^2}=-{2\over 3 \alpha}$.  To see angels and devils moving in the Poincar\'e disk click here: http://youtu.be/milmZUVSjro 
 }
\label{f3}
\end{center}
\vspace{-0.3cm}
\end{figure}

The hyperbolic plane ${ \mathbb{H}^2}$ has a long history in mathematics and physics, see for example \cite{Narayanan:1989uz}. A set of  
 user-friendly references with pictures and applications in physics include
\noindent http://mathworld.wolfram.com/PoincareHyperbolicDisk.html
https://www.youtube.com/watch?v=JkhuMvFQWz4

The Poincar\'e disk model of a hyperbolic geometry is presented  by the Escher's picture Circle Limit IV, see Fig.~\ref{f3}. The   boundary circle (which
is not part of the hyperbolic plane) is called the {\it absolute}. 
 One can place an infinite amount of angels and devils, of the  size which looks decreasing, towards the boundary in this circle, as Escher did.  However, in fact, the correct understanding of hyperbolic geometry means that the angels and devils close to the boundary are of the same `physical' size as the ones near the centrum of the circle. How do we explain this? As always in a curved space the concept of a distance (or size) depends on the geometry and there is a difference between the coordinate distance and physical distance.

The moduli space metric of these models, associated with the kinetic term of the scalars field is
\be
ds^2={1\over 2} {\partial \phi^2\over (1-{\phi^{2}\over 6\alpha})^{2}}=  {d r^2\over (1-{r^{2}\over 3\alpha})^{2}}  \ .
\label{relat}\ee
Here $r= \phi/\sqrt 2 $. It 
may be viewed as a slice at a fixed angular direction of the  2d metric of the Poincar\'e disk:
\be
ds^2= {d r^2 +r^2 d\theta^2\over (1-{r^{2}\over 3\alpha})^{2}} \, , \quad r^2< 3\alpha \ .
\label{relat2}\ee
A Poincar\'e disk is a space with a constant negative curvature ${\cal R}_{ \mathbb{H}^2}= -{2\over 3\alpha}$. At $\theta$=const
this is a slice of the Escher's picture, at fixed angular direction. Note that the physical distance on the hyperbolic disk is 
\be
d\rho =  {d r \over 1-{r^{2}\over 3\alpha}}  \ .
\ee
When $\rho\rightarrow \infty$,  $r\rightarrow 3\alpha$, towards the absolute, towards the boundary. When $\theta$ is not fixed (not stabilized by a dynamical mechanism during the cosmological evolution) the supergravity $\alpha$-attractor models  actually have a kinetic term for scalars presenting a Poincar\'e disk model of a hyperbolic geometry in eq. \rf{relat2}, as we will show in a more technical Sec. 5.

\begin{figure}[ht!]
\begin{center}
\includegraphics[width=7.2cm]{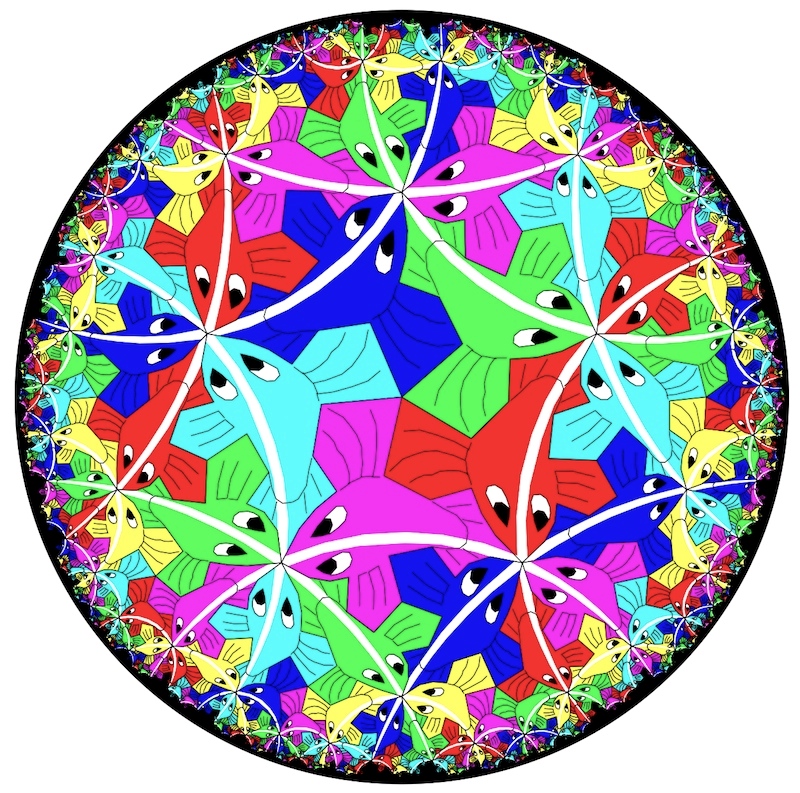}
\caption{\footnotesize A computer generated picture  inspired by
Escher's picture Circle Limit III http://www.math-art.eu/Documents/pdfs/Dunham.pdf. It presents a Poincar\'e disk model of a hyperbolic geometry. The symmetries of the geometry are shown here via configuration of fishes and how these configurations are mapped into other parts of a space. Mathematically, we will explain the symmetry in Sec.~7. }
\label{f4}
\end{center}
\end{figure}

At this point a cosmologist might  ask a question: why do we have to start with the complicated inflaton kinetic term shown in \rf{cosmo}, \rf{cosmoE} which we call here a moduli space? A simple answer to this question is: we have assumed that our $\alpha$-models have a certain symmetry, called 
M\"obius symmetry. It is a generic  symmetry of  superconformal theories. We show the picture associated with this symmetry,  in Fig. \ref{f4}, Escher's type picture of Circle Limit III.

\section{Relation to negatively curved 3-geometries in  FRW metric}

It is important to stress here that the metric of the moduli space in \rf{cosmo}, where the scalar fields  are coordinates of the manifold, is not a   metric of the space-time. In \rf{relat} we name the coordinate $r$ instead of $ \phi/\sqrt 2$ only for the purpose of inviting an intuition gained in general relativity with regard to space-time geometry, to be used for the geometry of the moduli space, where coordinates are scalar fields.\footnote{The importance of moduli spaces of a constant negative curvature in type IIA compactification of string theory for constructions of de Sitter vacua and cosmological inflationary models was stressed in \cite{Silverstein:2007ac}.}

In fact, the space of a constant negative curvature, which is a  Poincar\'e disk model of a hyperbolic geometry, reminds  the 3d slice at constant time of the familiar FWR geometry in case of the open universe with $k=-1$. It is  known that the 2d slice of the open FRW universe is related to Escher's picture Circle Limit IV, see for example the cosmology lecture by L. Susskind https://www.youtube.com/watch?v=H3D5HGZIP4s where after the 59 minutes into the class, the relation to Escher paintings is explained.  Observationally at present our 3d geometry is very close to flat with $k=0$. The corresponding parameter $\Omega_K$ is given by \cite{Planck:2015xua},
\be
\Omega_K= 0.000\pm 0.005 \ .
\ee
It appears that at present there is no indication that in our universe with the FWR model there is a negatively curved 3-geometry. However, more precise observations will take place in the future.

Meanwhile, the $\alpha$-models  suggest   a possibility of measuring the value of the curvature of a negatively curved 2-geometry not of a FWR model of a space and time but of the moduli space of scalars, which form a non-trivial geometry.

\section{Supergravity  $\alpha$-attractor models with  disk geometry}

Generic supergravity models are described by superfields and \K\, geometry. It means that the complex scalars are coordinates of the \K\, manifold. In the simplest case of the $\alpha$-attractor models we will focus only on the inflaton superfield $Z= z(x) +i a(x)$. The corresponding \K\, geometry can be described as a disk geometry defined by a \K\, potential of  the form $K= -3 \alpha \log  (1- Z\bar Z )$ where $Z\bar Z <1$. The moduli space metric is defined as 
\be
 ds^2=  g_{Z\bar Z}d Z d \bar Z\, ,    \qquad g_{Z\bar Z}= K_{Z\bar Z} = {3\alpha\over (1- Z\bar Z)^2} \ .
\label{met}\ee
Formally, we may proceed by a computation of the \K\, manifold curvature using the definition of it via the metric:
\be
 {\cal R}_{\rm K\ddot{a}hler}= - g_{Z\bar Z}^{-1} \partial_Z \partial_{\bar Z} \log g_{Z\bar Z}= -{2\over 3 \alpha} \ .
\label{curv}\ee
This is not quite illuminating, so we may try to do better using the Escher's Circle Limit picture.
A very nice  interpretation of $\alpha$ in \rf{met} and in \rf{curv}  comes from the concept of the Poincar\'e disk model of a hyperbolic geometry, as we pointed out around eq. \rf{relat}. First we establish that our disk \K\, geometry is actually a Poincar\'e disk model of a hyperbolic geometry
\be
  ds^2=  {3\alpha\over (1- Z\bar Z)^2} d Z d \bar Z = {d x^2+d y^2 \over \Big (1-{ x^2+ y^2\over 3\alpha}\Big )^{2}}    \, , \label{old Lag-alpha}
\ee
where $Z= ( x+i y)/\sqrt {3\alpha}$. We have shown this geometry in polar coordinates in eq. \rf{relat2}.
The physical distance in this geometry is defined by 
$
d\rho =  {d r \over 1-{r^{2}\over 3\alpha}} \ ,
$
and we find that 
\be
r= \sqrt{3\alpha} \tanh{\rho\over\sqrt {3 \alpha}} \ .
\ee
When $\rho \rightarrow \infty$, $r$ never reaches $3\alpha$ since $\tanh  <1$.
The curvature of the Poincar\'e disk of a radius $R= \sqrt {3\alpha} $ is equal to $-2/R^2$
\be
{\cal R}_{\rm Poincar\acute{e}} =  -{2\over 3 \alpha} \ .
\ee
Finally, we can convert our geometry into a well known metric of an open 2d universe with a negative curvature $R_{\rm open} =   -{2\over 3 \alpha}$
\be
 ds^2=   {3\alpha\over (1- Z\bar Z)^2} d Z d \bar Z =  {3\alpha\over 4}  (d\chi^2 + \sinh^2 \chi d\theta^2) \ ,
\label{open} \ee
 where the following change of variables was performed
\be
Z=  \, e^{i\theta} \, \tanh {\chi\over 2}\ .
\label{change}\ee
Note that the relation between $\chi$ and a canonical field $\vp$ is given by  
\be
{\chi\over 2} = {\vp \over \sqrt {6\alpha}} \ .
\ee

\begin{figure}[ht!]
\begin{center}
\includegraphics[width=7 cm]{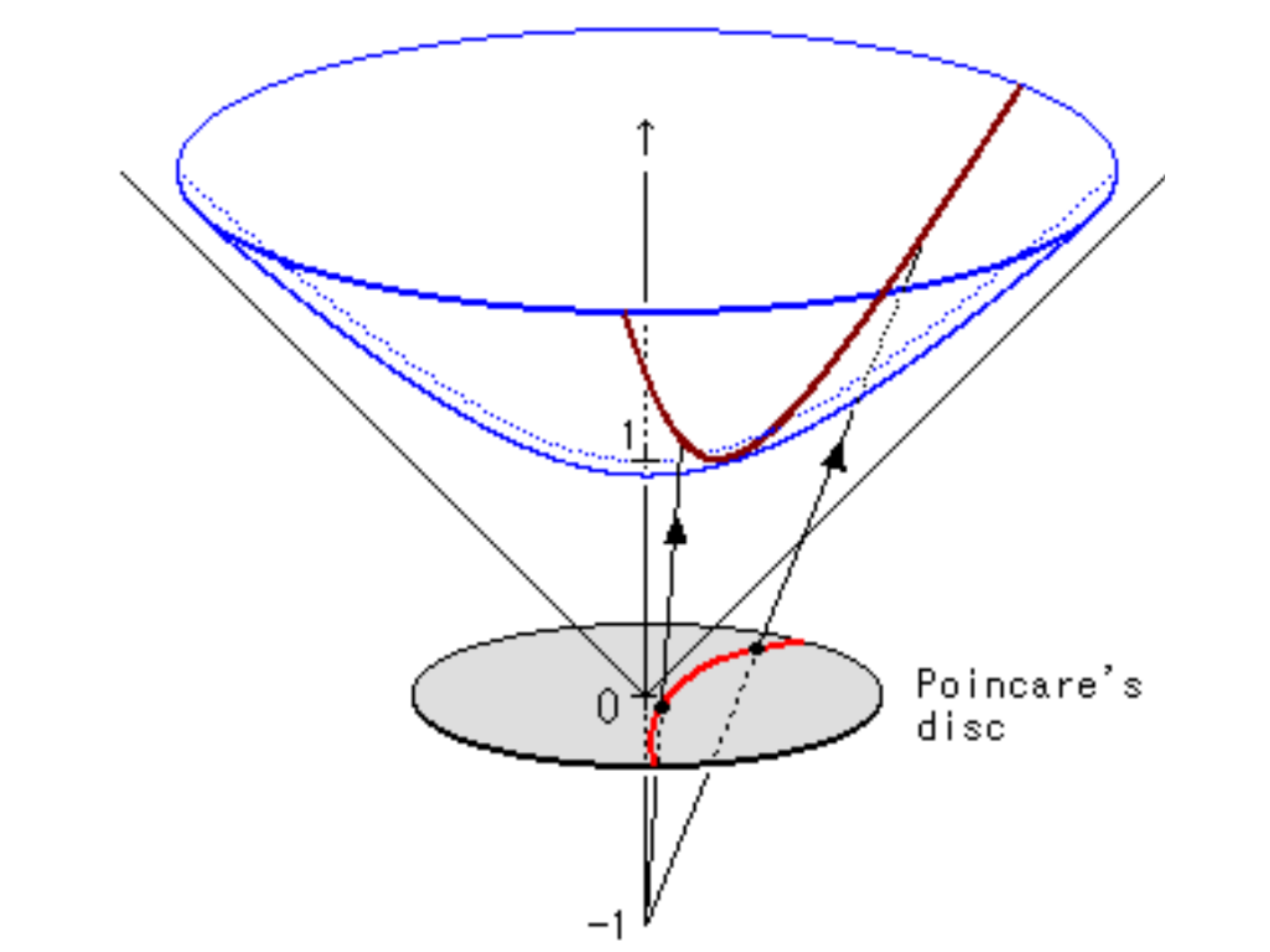}
\vspace*{-0.4cm}
\caption{\footnotesize A unit size hyperboloid given in eq. \rf{hyper}, see Wikipedia for Poincar\'e disk model.   In this picture the
Poincar\'e disk model is a perspective projection viewed from the point $w=- 1, u=v=0$ projecting the upper half hyperboloid onto an $u,v$ disk of  at $w=0$. The red circular arc is geodesic in Poincar\'e disk model; it projects to the brown geodesic on the blue hyperboloid. The figure shows the Poincar\'e disk of a radius $R=\sqrt {3\alpha}= 1$.
}
\label{f5}
\end{center}
\vspace{-0.3cm}
\end{figure}

\begin{figure}[ht!]
\begin{center}
\includegraphics[width=9.8 cm]{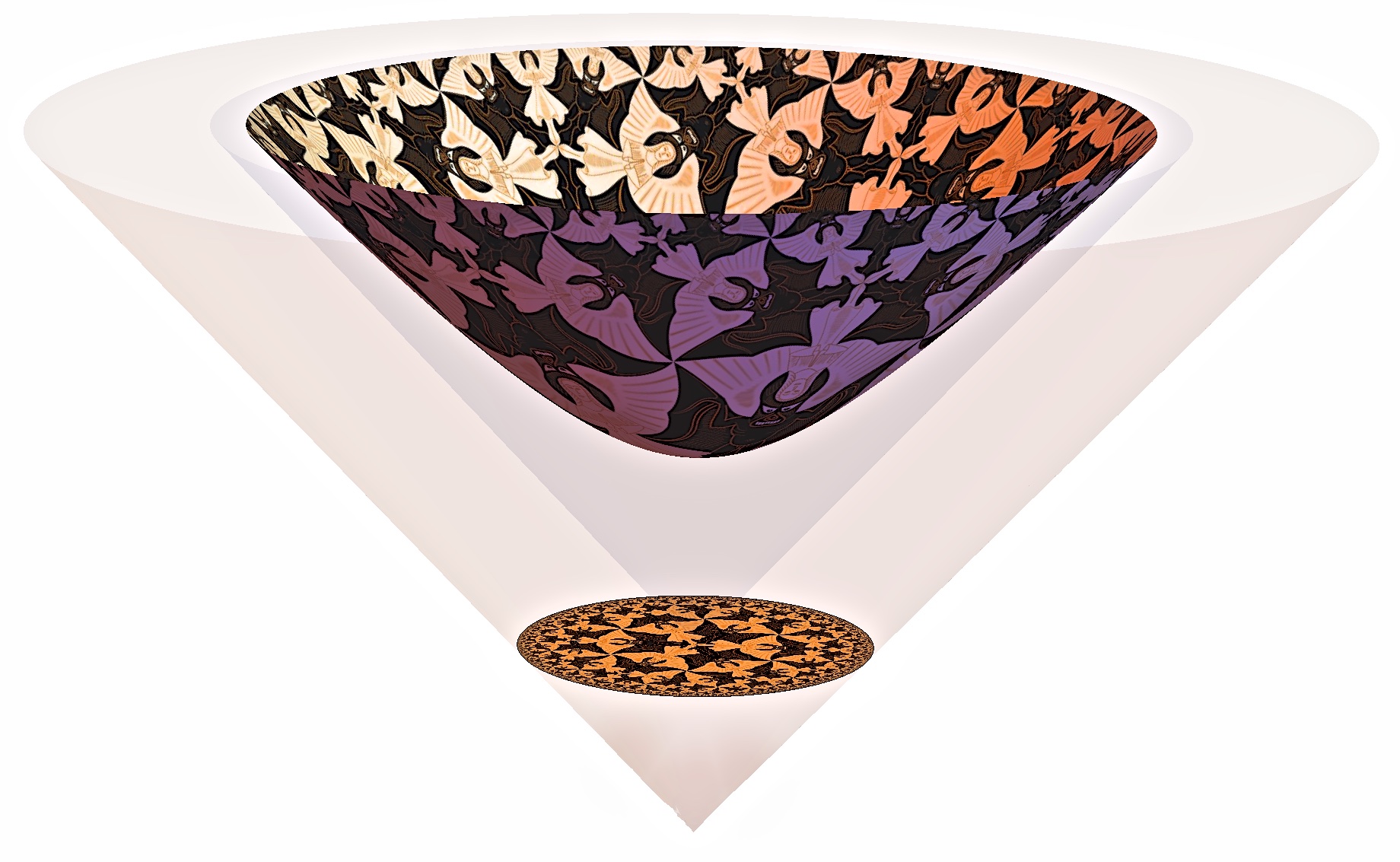}
\vspace*{-0.4cm}
\caption{\footnotesize An artistic version of Fig. \ref{f5}, which shows that angels and devils crowded near the boundary of the Poincar\'e disk have an open space on the hyperboloid, when projected from the disk.}
\label{f6}
\end{center}
\vspace{-0.3cm}
\end{figure}

A complementary point of view on this geometry  is given by a Minkowski metric in the embedding 3d space
\be
ds^2= {1\over 4} (du^2+ dv^2-dw^2) \ ,
\ee
 where the coordinates are restricted to a hyperboloid. This hyperboloid is associated with the negative space curvature (and an open universe, when the geometry if a part of FRW metric).
 \be
 -u^2 - v^2 + w^2 = 3\alpha  \ .
\label{hyper} \ee
  If we resolve this condition by taking 
  \bea
u&= & \sqrt {3\alpha} \sinh \chi  \cos \theta \nonumber \\
 v&= & \sqrt {3\alpha} \sinh \chi \ \sin \theta \nonumber \\
w&= & \sqrt {3\alpha} \cosh \chi 
  \eea
 with $0\leq \chi < \infty$ and $0\leq \theta\leq 2 \pi$
  we recover the geometry in eq. \rf{open} and the upper part of the hyperboloid in Fig. \ref{f5}. An artistic version of it is shown in Fig. \ref{f6}.
In our case we have a geometry based on one complex scalar, which plays a role of coordinates in the moduli space. 

\section{Superconformal  $\alpha$-attractor models: Escher's Circle and Escher's Half-Plane}

\begin{figure}[ht!]
\vspace*{0.2cm}
\begin{center}
\includegraphics[width=8.5 cm]{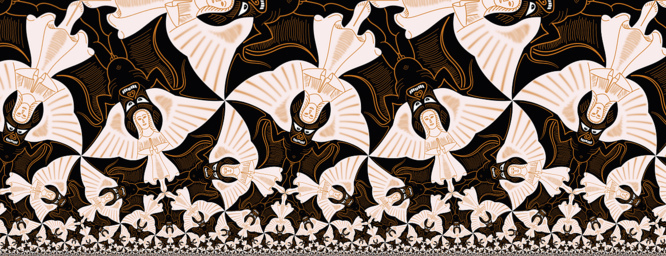}
\vspace*{0.1cm}
\caption{\footnotesize Escher's picture of a Heaven and Hell in half-plane variables. The  boundary of the half-plane $T+\bar T\rightarrow 0$ is
 the absolute which cannot  be reached. The angels and devils look smaller and smaller near the boundary. 
 }
\label{f7}
\end{center}
\end{figure}

\begin{figure}[ht!]
\begin{center}
\includegraphics[width=6 cm]{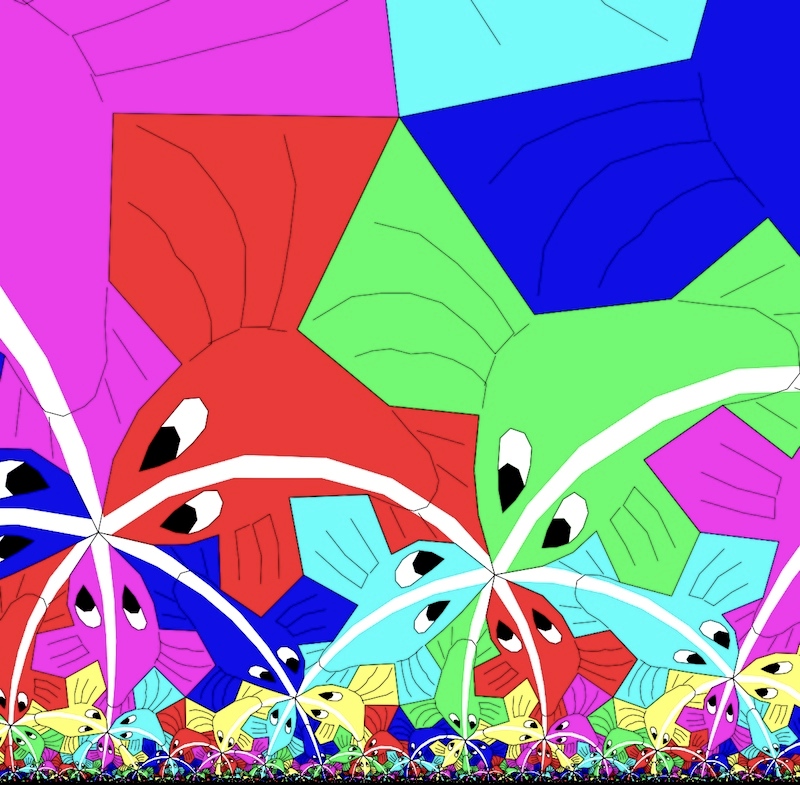}
\vspace*{0.1cm}
\caption{\footnotesize  A computer generated picture by D. Dunham http://www.math-art.eu/Documents/pdfs/Dunham.pdf, representing a half-plane geometry version of Fig. \ref{f4}.
 }
\label{f8}
\end{center}
\vspace{-0.3cm}
\end{figure}

The disk geometry is described by a complex variable $Z$ such that $Z\bar Z<1$. A    half plane geometry in $T$ variables with $T+\bar T>0$ is related to it by a change of variables \cite{Cecotti:2014ipa} 
\be
Z= {T-1\over T+1}\, ,  \qquad T= {1+Z\over 1- Z} \ .
\label{Galey}\ee
The corresponding disk geometry is presented by Escher's Circle Limit IV in Fig. \ref{f3}, and a Half-Plane one by an Escher's Half-Plane in Fig. \ref{f7}. Same for Escher's Circle Limit III in Fig. \ref{f5}, and a Half-Plane one  in Fig. \ref{f8}. 

The analysis of cosmological $\alpha$-models of inflation in disk and half-plane variables was performed in \cite{Kallosh:2013yoa,Cecotti:2014ipa}. The  generalized cosmological $\alpha$ models of inflation and dark energy with susy breaking, in disk and half-plane variables, were introduced  in \cite{Kallosh:2015lwa}. A detailed  analysis of 
 stability  of these models   will be presented in \cite{JJ}.

The \K\, potential in half-plain coordinates is $K= -3 \alpha \log \Big (T+\bar T)$. 
The curvature of the \K\, manifold is computed using the \K\, metric $ ds^2=  g_{T\bar T}d T d \bar T$ with   $g_{T\bar T}= K_{T\bar T} = {3\alpha\over (T+\bar T)^2}$:  
\be
 {\cal R}_{\rm K\ddot{a}hler}= - g_{T\bar T}^{-1} \partial_T \partial_{\bar T} \log g_{T\bar T}= -{2\over 3 \alpha} \ .
\ee
Since the relation between the disk and half-plane is due to a change of coordinates $Z= {T-1\over T+1}$, it is not surprising that the curvature in the half-plain coordinates is the same as in the disk ones. 

 An alternative form of this negative constant curvature space associated with Figs. \ref{f7}, \ref{f8}  can be also given in terms of the constant scalar curvature metrics on toric manifolds  \cite{Abreu}. The  scalar curvature for the metrics on toric manifolds  is
\be
 ds^2= { 3\alpha \over y^2} dy^2  \quad \Rightarrow   \quad {\cal R}_{\rm toric}= - \Big [ {1\over g(y)} \Big ]^{''}=  -{2\over 3 \alpha} \ ,
\label{toric1}\ee
where $u(y)$ is its symplectic potential and $g= u''$.

In all cases we find the same result for the curvature: ${\cal R}=-{2\over 3 \alpha}$, but the  Escher's circle limit pictures in Fig. \ref{f3} and  Fig. \ref{f5} help us to 
provide a simple interpretation of the parameter $\alpha$.  It will eventually be measured (or  bounded) in the context of the $\alpha$-attractor models, by looking at the primordial gravity waves from the sky.

The origin of the Poincar\'e disk model and of a half-plane geometry in supergravity models of the $\alpha$-attractors can be traced back to studies of ${\cal N} =4$ supergravity in \cite{Cremmer:1977tt}, where  it was shown that the disk action for the moduli ${ \partial  Z \partial  \bar Z\over (1- Z\bar Z)^2} $ has an $SU(1,1)$ symmetry. They have also explained that   for  ${\cal N} =1$ supergravity  a more general class of models with
$  3\alpha{ \partial  Z \partial  \bar Z\over (1- Z\bar Z)^2} $ still has an $SU(1,1)$ symmetry
for an arbitrary $\alpha$. The same symmetry was also discovered in a maximal superconformal model \cite{Bergshoeff:1980is}  which has a local $SU(4)\times U(1)$ symmetry and a global $SU(1,1)$ symmetry. Upon gauge-fixing some of the superconformal symmetries it becomes an ${\cal N} =4$ supergravity with the remaining $SU(1,1)$  symmetry. With one choice of a gauge \cite{Bergshoeff:1980is} one finds a disk geometry with $Z$-variables, $Z\bar Z <1$, the absolute is given by equation $Z\bar Z =1$.
With the other choice one finds a geometry of the half-plane, with $T$-variables, $T+ \bar T >0$, the  boundary is at   $T+\bar T=0$.
See \cite{Ferrara:2012ui}, eqs. (35) and (39) there, explaining  both choices. These two choices  are shown 
 Limit Circles  in Figs.  \ref{f3} and \ref{f5} and in Half-Plane ones  in Figs. \ref{f7}, \ref{f8}.  

The relation between these two geometries corresponds to a different choice of a local $U(1)$ $\mathbb{R}$-symmetry gauge in the superconformal theory,  \cite{Ferrara:2012ui}. On the other hand, a simple change of variables preserving the geometry $ds^2$ was explained for our attractor models in \cite{Cecotti:2014ipa} via a Cayley transform \rf{Galey}.

An origin of {\it manifolds with boundaries} can be traced also to the coset space structure of extended supergravities. For example in maximal ${\cal N} =8$ supergravity, with scalars in the coset space ${E_{7,7} \over SU(8)}$, the positivity of kinetic terms of these scalars requires a condition  \cite{Cremmer:1979up}, which upon truncation to ${\cal N} =4$ supergravity becomes $Z\bar Z <1$ with $3\alpha=1$ .

 We may start with maximal ${\cal N}=4$ superconformal model and gauge-fix some local symmetries, including the Weyl symmetry  \cite{Bergshoeff:1980is,Ferrara:2012ui}, or perform a supersymmetric truncation of the maximal ${\cal N}=8$ supergravity \cite{Cremmer:1979up}. For pure ${\cal N}=4$ supergravity we recover in both cases the following  bosonic action, see for example eq. (A.1) in \cite{Cremmer:1979up}:
 \be
 {1\over \sqrt{-g}} \mathcal{L}_{{\cal N}4} = {1\over 2}   R - {d Z d \bar Z \over (1- Z\bar Z)^2}  +{1\over 4} F_{\mu\nu} ^{ab}  F^{\mu\nu cd} {\cal M}_{abcd} (Z, \bar Z)\,  .
\label{N4}\ee
There is no potential, the kinetic term of the scalar $Z$ represents an unit radius Escher disk geometry, the scalars interact with vectors $F_{\mu\nu} ^{ab}$. For inflationary period (but not for the reheating stage) we may ignore vectors. If we were to associate this bosonic model with ${\cal N}=1$ supergravity, we would qualify it as $K=- 3\alpha \ln (1- Z\bar Z)$ with $\alpha=1/3$ and $W=0$.

\section{Isometries of the half-plane and the disk geometries}

Here we focus on symmetries of our geometries. It involves the  $GL(2, \mathbb{R})$ M\"obius transform.   A simple form of it  is given in terms of   $\tau$ variables, familiar to a string theorist, where  $\tau = i T$. Namely, our half-plane geometry is given by
\be
 ds^2=  3\alpha {d T d \bar T \over (T+\bar T)^2} = 3 \alpha {d\tau d \bar \tau \over (2 \, {\rm Im} \tau)^2} \ .
 \ee
It is invariant 
under the  transformations $\tau\rightarrow \tau'$, where
  \be
\tau'= {a\tau+b\over c\tau +d},  \qquad ad-bc\neq 0 \ ,
\label{sl}\ee
where $a,b,c,d$ are real numbers and 
\be
{ d \tau d \bar \tau\over (\tau-\bar \tau)^2}=  { d \tau' d \bar \tau'\over (\tau'-\bar \tau')^2}\ .     
\ee

Note that the $GL(2, \mathbb{R})$ isometry of the half-plane is valid for  $ad-bc\neq 0$ and does not require that $ad-bc= 1$, which corresponds to an $SL(2, \mathbb{R})$ symmetry. Thus the isometry is a  general linear group over $\mathbb{R}$ and includes a special linear group over $\mathbb{R}$.
The symmetry is valid for any $\alpha$, it defines the geometry of the moduli space which we are employing. 

An analogous symmetry acts on the disk geometry. This is a  M\"obius transform of a Poincar\'e disk.
  \be
Z'= {\beta Z+\gamma\over \bar \gamma Z +\bar \beta},  \qquad |\beta|^2 - |\gamma|^2 > 0 \ .
\label{sl2}\ee
Here $\beta$ and $\gamma$ are complex numbers.
The symmetry is the same, the properties of the geometry are the same for all $\alpha$, however, the size of the Escher's Limit Circle is different, its radius square is $3 \alpha$. 
In the context of our $\alpha$-attractor models we will measure $\alpha$  when the primordial gravity waves will be discovered. 

 Some special choices for $\alpha$ are: $\alpha=1/3$ and a unit size Escher disk  $R=1$ correspond to a maximal ${\cal N}=4$ superconformal model and pure ${\cal N}=4$ supergravity, $r\sim 10^{-3}$. The case $\alpha=1$ and an Escher disk $R=3$ support the geometry of the ${\cal N}=1$ superconformal model, $r\sim 3\times 10^{-3}$. Finally, ${\cal N}=1$ supergravity geometry is consistent with an arbitrary positive $\alpha$.
 
 \section{From moduli space  to cosmology}
 
 Until now, we focused on the geometry of the moduli space, described by the kinetic term in the Lagrangian. There was a good reason to do it: Once we decide on a potential, we can study the evolution of the observable universe, and compare it with the data in \cite{Ade:2015tva,Planck:2015xua}, and especially with the future data. There are many options with regard to the choice of a potential.

 A generic class of inflationary models \cite{Kallosh:2015lwa} compatible with the current data, as well as capable of describing dark energy and controllable susy breaking, involve an addition chiral superfield $S$, which can be arranged to vanish during and after inflation.  The corresponding \K\, potential  is now $K=- 3\alpha \ln (1- Z\bar Z - {S\bar S})$
and in the context of  ${\cal N}=1$ supergravity we can make a choice of a holomorphic superpotential $W= A(Z) + SB(Z)$.

A very simple choice here comes from the ${\cal N}=4$ model \rf{N4} which suggests to use $3\alpha=1$. We also take a very simple superpotential $A(Z)=0$ and $B(Z)=\mu$:
\be
K=-  \ln (1- Z\bar Z - {S\bar S})\, ,  \quad W= \mu \, S \ .
\label{KW}\ee 
 This leads to the theory with the bosonic  action
  \be
 {1\over \sqrt{-g}} \mathcal{L}_{{\cal N}4 \rightarrow  {\cal N}1} = {1\over 2}   R - {d Z d \bar Z \over (1- Z\bar Z)^2}  - \mu^2\,  .
\label{N41}\ee
This bosonic model has  an embedding into ${\cal N}=1$ supergravity, according to \rf{KW}. It also has an unbroken  the M\"obius symmetry \rf{sl2}. Its moduli space is the Poincar\'e disk with unit radius $R= 1$. And, from the point of view of cosmology, it describes de Sitter space with a positive vacuum energy $V = \mu^{2}$ and spontaneously broken ${\cal N}=1$ supersymmetry.
Thus we are coming very close to describing inflation. We have dS space, but now we must find a way to end the stage of the exponential expansion in dS vacuum.

\begin{figure}[ht!]
\begin{center}
\includegraphics[width=9.5 cm]{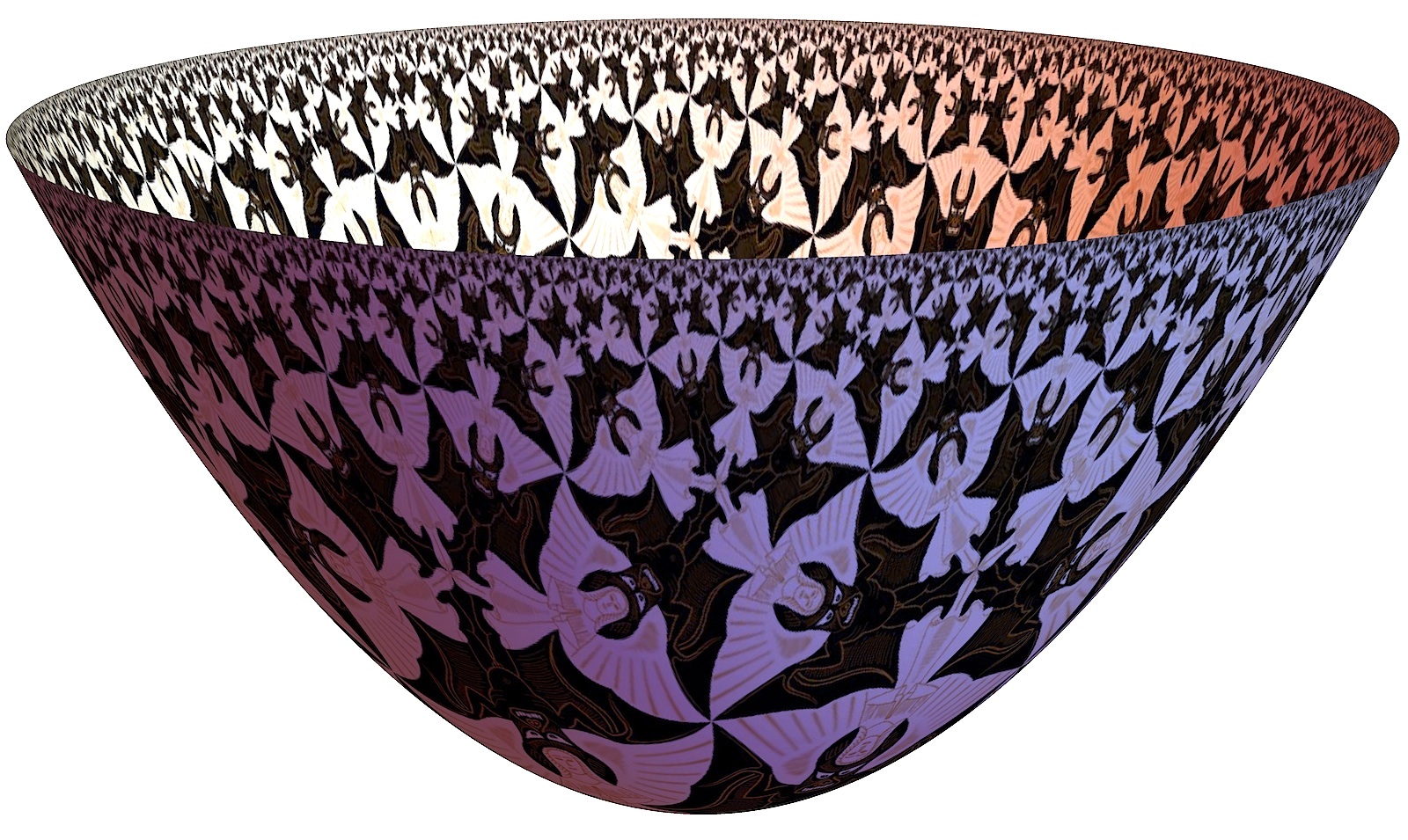}
\vspace*{-0.4cm}
\caption{\footnotesize  The simplest quadratic inflationary potential $V(Z, \bar Z)= \mu^2 Z \bar Z$
in the theory \rf{N41s}. The picture reveals the fact that the potential depends on disk coordinates with $Z \bar Z<1$ and the angels and devils are crowded near the top of the potential close to the boundary at $|Z|=1$.
}
\label{cup}
\end{center}
\vspace{-0.3cm}
\end{figure}

As a next step, we consider the same model but with the superpotential 
\be
W= \mu \, S Z. 
\ee
This brings the action to the form closely resembling the simplest toy model  \rf{cosmo} we started with:
  \be
 {1\over \sqrt{-g}} \mathcal{L}_{{\cal N}4 \rightarrow  {\cal N}1} = {1\over 2}   R - {d Z d \bar Z \over (1- Z\bar Z)^2}  - \mu^2\, Z\bar Z .
\label{N41s}\ee
This model has a simple quadratic potential with respect to the complex field $Z$. However, in the theory \rf{N41} the value of the potential was everywhere the same across the  Poincar\'e disk, whereas in \rf{N41s} it approaches its maximum value close to the boundary of the moduli space at $|Z| = 1$, as shown in Fig. \ref{cup}. 
As one can easily see, most of the angels and devils live close to this boundary. It could seem that they do not have much space here, and they should quickly fall down instead of hanging up in the sky. But this is not the case. 

\begin{figure}[ht!]
\begin{center}
\includegraphics[width=9.5 cm]{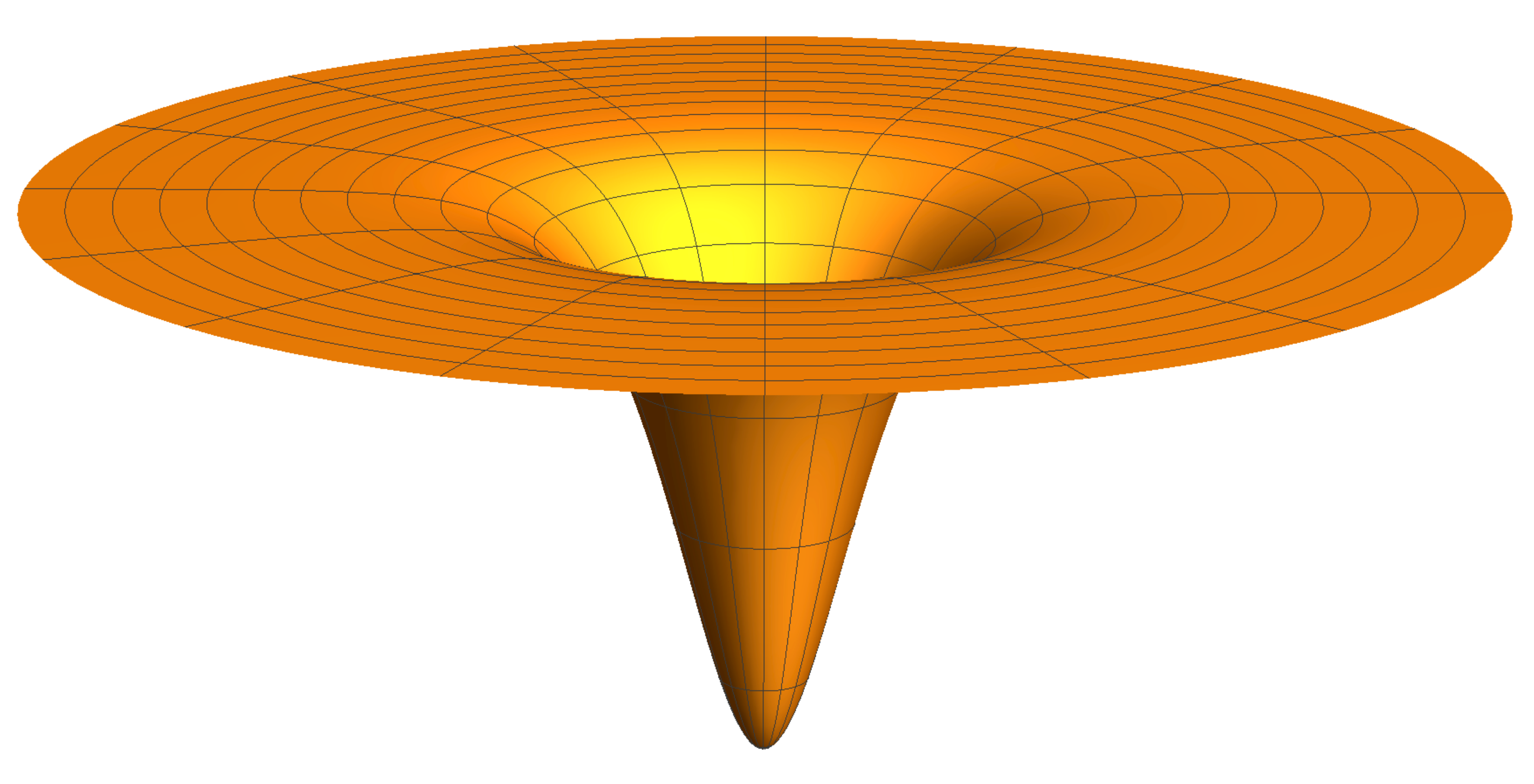}
\vspace*{-0.4cm}
\caption{\footnotesize Inflationary potential $V$ in the theory \rf{N41s}, \rf{N41can} with the radial direction represented by the canonical field $\vp$ with $V\sim  \tanh^2 {\vp\over \sqrt2}$.}
\label{conf}
\end{center}
\vspace{-0.5cm}
\end{figure}

Indeed, if we represent the radial component in terms of the canonical field $\vp$,  as in eq. \rf{change}, $
Z=  e^{i\theta} \, \tanh {\vp\over 2},
$ our action \rf{N41s} becomes
 \be
{R\over 2} - {1\over 2} \left(d\vp^2 + {1\over 2} \sinh^2(\sqrt 2 \vp)\,  d\theta^2\right) -\mu^2 \tanh^2 {\vp\over \sqrt 2} \ .
\label{N41can}\ee
When $|Z|$ approaches the boundary of the moduli space, the canonical field $\vp$ runs to infinity. This means, in effect, that the upper part of the paraboloid shown in Fig.~\ref{cup} becomes infinitely stretched out, and this part of the potential becomes exponentially flat, as shown in Fig.~\ref{conf}, with the height of the plateau asymptotically approaching the vacuum energy $\mu^{2}$ in dS space described by \rf{N41}. 

Fig.~\ref{conf} does not give full justice to the volume of the moduli space at the plateau of the potential, because with the growth of the canonical field $\vp$, shown by the circles, the distance in the angular direction grows exponentially fast, as $\sinh(\sqrt 2 \vp)$, see  \rf{N41can}. Therefore most of this volume is at indefinitely large values of the field $\vp$. In other words, in accordance with this model, almost all angels and demons live at this high plateau. 

This provides perfect initial conditions for inflation: Independently of the initial velocity of the scalar fields, their kinetic energy rapidly dissipates due to the cosmological evolution, the fields freeze at some point of the infinitely large plateau, until the exponentially slow descent in the radial direction towards the minimum of the potential shown in Fig. \ref{conf} begins. Unless one makes an unusual assumption that the decay rates of the radial and angular components of the field are dramatically different from each other, the perturbations of metric produced during inflation in this theory have the same properties as the perturbations in the simplest single-field model \rf{cosmo} with $\alpha = 1/3$, which perfectly match the recent cosmological data, as shown  in Fig.~\ref{f1}. Moreover, one can consider a more general superpotential, 
\be
W = \mu S\, f(Z), 
\ee
obtain a more general potential 
\be
V= \mu^2\,  f^2(\tanh {\vp\over \sqrt 2}), 
\ee
and show that for a very broad choice of the functions $f(Z)$ this theory has the same observational predictions 
\cite{Kallosh:2013hoa,Kallosh:2013yoa}. 

One may also consider the \K\ potential for general $\alpha$, 
\be
K=-  3\alpha \ln (1- Z\bar Z - {S\bar S}), 
\ee 
and perform the same two-step procedure: construct dS space, and then deform it in the place corresponding to the end of inflation, e.g. at $Z = 0$. In order to do it, one may consider superpotentials 
\be
W=  \mu \, S (1-Z^{2})^{3\alpha-1\over 2}.
\ee
The geometry of the moduli space will be described by the Poincar\'e disk of the radius $3\alpha$. The potential of the field $Z$ will be given by the cosmological constant $\mu^{2}$, but  only for real values of the field $Z$, i.e. for $\theta = 0$. The potential has a stable minimum with respect to $\theta$ at $\theta =0$ for $\alpha > 1/3$. In this case, the field $\theta $ is frozen but the field $\phi$ is free to move, so we can proceed the same way as before. Multiplying the superpotential by a function  $f(Z)$, so that 
\be
W= \mu \, S f(Z) (1-Z^{2})^{3\alpha-1\over 2} \ ,
\ee
 results in a theory with the  inflaton potential 
 \be
 V=    \mu^2\,  f^2(\tanh{\vp\over \sqrt{6\alpha}}) \ . 
 \ee
Thus we recover a broad class of the single-field T-models and E-models discussed in Section~\ref{single}.

Other examples of cosmological $\alpha$-attractors include supergravity models with a single inflaton field $\vp$ \cite{Ferrara:2013rsa,Goncharov:1983mw,Kallosh:2015lwa,Roest:2015qya,Linde:2015uga}.
One may also modify the two-field models described above by considering different \K\ potentials and superpotentials, allowing stable inflation for $\alpha < 1/3$, breaking supersymmetry spontaneously and uplifting the minimum of the potential to account for the tiny vacuum energy (cosmological constant) $V_{0}\sim 10^{-120}$, without altering the main cosmological predictions of $\alpha$-attractors \cite{Kallosh:2015lwa,JJ}. 

This stability of the predictions with respect to even very significant changes of the potential is the main reason why we called these theories ``cosmological attractors:'' Their predictions for $n_{s}$ and $r$ are mostly determined by the underlying geometry of the moduli space rather than by the choice of the inflaton potential. That is why the knowledge of the geometry of the moduli space may be important for cosmology, even if the initial symmetry of the theory is hidden from us by spontaneous supersymmetry breaking and by the structure of the potential. This suggests that cosmological observations 
may help us to explore the geometric structure of the theory of all fundamental interactions.

\newpage

\subsection*{Acknowledgements}

Maurits Cornelis Escher  was a Dutch graphic artist. Some of his works were inspired by advanced developments in mathematics, some other works featured impossible constructions which tricked and misled our eye and imagination, and some were a combination of both.  We are grateful to R. Bond,  L. Page and  U. Seljak for asking us about the meaning of the $\alpha$-parameter in our models, and for not being fully satisfied by our formal answer that it is inversely proportional to the curvature of the \K\ manifold. We are grateful to our collaborators  S. Ferrara, M. Porrati and D. Roest  for developing these cosmological models and to E. Bergshoeff, P. Binetruy, F. Bouchet, J.J. Carrasco, M. Gunaydin, S. Kachru, L. Senatore, E. Silverstein, A. Strominger, L. Susskind and A. Van Proeyen for many stimulating discussions.
 RK and AL are supported by the SITP and by the NSF Grant PHY-1316699. RK is also supported by the Templeton foundation grant `Quantum Gravity Frontiers,' and AL is supported by the Templeton foundation grant `Inflation, the Multiverse, and Holography.'

\end{document}